\documentclass[preprint,preprintnumbers,nofootinbib,amsmath,amssymb]{revtex4}

\usepackage{amsmath}
\usepackage{graphicx}
\usepackage{dcolumn}
\usepackage{bm}

\begin{document}


\title{Reconstruction of the Primordial Power Spectrum by Direct Inversion}

\author{Gavin Nicholson}
 \email{gavin.nicholson05@imperial.ac.uk}
\author{Carlo R. Contaldi}%
 \email{c.contaldi@imperial.ac.uk}
\affiliation{%
Theoretical Physics, Blackett Laboratory, Imperial College, Prince
 Consort Road, London, SW7 2BZ, U.K.}%
\author{Paniez Paykari}%
 \email{p.paykari06@imperial.ac.uk}
\affiliation{%
Astrophysics Group, Blackett Laboratory, Imperial College, Prince
 Consort Road, London, SW7 2BZ, U.K.}%

\date{\today}

\begin{abstract}
  We introduce a new method for reconstructing the primordial power
  spectrum, $P(k)$, directly from observations of the Cosmic Microwave
  Background (CMB). We employ Singular Value Decomposition (SVD) to
  invert the radiation perturbation transfer function. The degeneracy
  of the multipole $\ell$ to wavenumber $k$ linear mapping is thus
  reduced. This enables the inversion to be carried out at each point
  along a Monte Carlo Markov Chain (MCMC) exploration of the combined
  $P(k)$ and cosmological parameter space. We present best--fit $P(k)$
  obtained with this method along with other cosmological parameters.
\end{abstract}

\maketitle

\section{Introduction}

The primordial power spectrum of scalar, curvature perturbations
$\Phi({\vec k})$ is defined as,
\begin{equation}\label{eq:pkdef}
P(k) \equiv \frac{k^3}{2\pi^2}\delta^3(\vec{k}-\vec{k'})\langle\Phi(\vec{k})\Phi^*(\vec{k'})\rangle , 
\end{equation}
where $k\equiv |\vec k|$ is the wavenumber. The spectrum encodes the
initial conditions for the system of coupled Einstein--Boltzmann
equations which describe the evolution of density and radiation
perturbations about the FRW background. The spectrum itself is
considered as a unique window into the era approaching the
Planck time. In most models a period of scalar field driven inflation
is used to solve the cosmological problems and set the near
scale-invariant form of $P(k)$ prior to the radiation dominated
epoch. In this case a simple power law parametrised by an amplitude
$A_s$ and spectral index $n_s$ suffices to describe the initial
conditions to sufficient accuracy for the current data. Higher
order contributions such as a mild curvature $d n_s/d\ln k$  have also
been explored although it is not strictly required by the best--fit
models. 

A separate approach to studying the physics behind this early phase is
to drop any model dependent assumptions of near scale invariance and
allow more general functional forms of the initial spectrum. One
drawback of this approach is the increase in parameter space which
needs to be explored which increases the complexity of the data
fitting step. A second drawback is the limited information content of
the observations; the effect of sample variance, limited range of
scales probed, and the degeneracy in the mapping of $k$ to angular
multipoles on the sky means that one cannot expect to constrain
arbitrarily complex functions with many degrees of freedom. However
there is scope to go beyond the model dependent approach. This is
particularly true if one is interested in constraining the presence of
features on the spectrum. These could be in the form of `glitches' or
step--like features which would be otherwise unconstrained. In fact
many models have been proposed which predict features on the spectrum
due to, for example, features on the inflaton potential
\cite{Starobinsky:1992ts,Adams:2001vc,Wang:2002hf,Hunt:2004vt,Joy:2007na,Hunt:2007dn,Pahud:2008ae,Lerner:2008ad},
a small number of $e$-folds
\cite{Contaldi:2003zv,Powell:2006yg,Nicholson:2007by}, or other more exotic
sources of non-standard behaviour 
\cite{Lesgourgues:1999uc,Feng:2003zua,Mathews:2004vu,Jain:2008dw,Romano:2008rr}.



There are two approaches to reconstructing $P(k)$; parameterisation
and direct inversion. None of the various methods have shown
conclusive evidence for a departure from near scale-invariance of
$P(k)$. Despite this there have been tantalising hints of anomalous
features in the CMB. One example of this is that the first year WMAP
results gave an indication of a cut-off in $P(k)$ on large
scales. With subsequent data releases the significance of this feature
has been reduced, although future observations of the polarisation of
the CMB may provide more conclusive evidence
\cite{Nicholson:2007by,Jain:2009pm}. In \cite{Nicholson:2009pi} we
also showed evidence for a dip in power at $k\approx0.002$Mpc$^{-1}$.

Numerous parametertic searches for features with a similar form to
those in complex inflationary models have been performed along with
simple binning of $P(k)$
\cite{Bridle:2003sa,Contaldi:2003zv,Parkinson:2004yx,Sinha:2005mn,Sealfon:2005em,Mukherjee:2005dc,Bridges:2005br,Bridges:2006zm,Covi:2006ci,Joy:2008qd,Verde:2008zz}. Methods
of direct inversion which make no assumptions about the early universe
model being probed
\cite{Hannestad:2000pm,Wang:2000js,Matsumiya:2001xj,Shafieloo:2003gf,Bridle:2003sa,Kogo:2003yb,Mukherjee:2003cz,Mukherjee:2003ag,Hannestad:2003zs,Kogo:2004vt,TocchiniValentini:2004ht,Leach:2005av,Shafieloo:2006hs,Shafieloo:2007tk,Nagata:2008tk,Nagata:2008zj,Nicholson:2009pi}
are hampered by the singular nature of the transfer function that
takes $P(k)$ and transfers it onto the CMB or LSS modes. In general
this causes the process of estimation to be prohibitively slow so as
not to allow joint estimation of a free $P(k)$ with cosmological
parameters. Instead one usually assumes a set of cosmological parameters, this allows the use of a fiducial CMB photon transfer function to integrate the primordial curvature perturbation into today's photon distribution perturbation. However, this hides a significant degeneracy between features in the primordial power spectrum, and the physical parameters which determine the height and position of acoustic peaks in the CMB. It has been pointed out
\cite{Hu:2003vp,Nicholson:2009pi,Mortonson:2009qv} that adding
polarisation information or LSS data can help break some of the
degeneracies.

In this paper we propose a new method for direct inversion of $P(k)$
from CMB observation using Singular Value Decomposition (SVD). The
method is fast enough to allow us to carry out a joint estimation of
$P(k)$ and the cosmological parameters. The form of $P(k)$ is derived
from the SVD inversion. We also show that there are regions of $k$ for
which polarisation data has the potential to more accurately constrain
$P(k)$. The paper is organised as follows; in section~\ref{method} we
introduce the SVD method and test it against known input models in
section~\ref{results}. We also show current constraints from the joint
estimation of cosmological parameters and $P(k)$ reconstructions in
section~\ref{current}. We discuss our results and conclude in
section~\ref{discussion}.

\section{Direct Inversion of $P_k$ by Singular Value Decomposition}\label{method}

Direct primordial power spectrum reconstruction requires the inversion
of the following relations
\begin{equation}\label{eq:intro-aps}
C_\ell^{XY} =\int\limits_{0}^{\infty} \frac{\rm{d}k}{k} \Delta_{\ell}^{X}(k) \Delta_{\ell}^{Y}(k) P(k),
\end{equation}
where $X$ and $Y$ represent $T$, $E$, or $B$-type anisotropies,
$C_\ell^{XY}$ are the angular power spectra for the $XY$ combination
and the $\Delta^{X}_{\ell}(k)$ are the photon perturbation transfer
functions. The transfer functions are obtained by integrating the full
Einstein-Boltzmann system of differential equations
\cite{cmbfast,camb}. These describe the evolution of perturbations in
the photon distribution functions in the presence of gravity and other
sources of stress-energy. The functions determine all of the structure
in the anisotropy spectra which arise after the initial conditions
are set. Most notably the $C_\ell^{XY}$ contain distinct peaks due to
the acoustic oscillation of the tightly coupled photon-baryon fluid in
gravitational potential wells at the time of last scattering. The aim
of any inversion method is to distinguish such features from any
structure in the initial perturbation spectrum.

For a finite sampling of the wavenumber space $k$
Eq.~(\ref{eq:intro-aps}) can be recast as an operator acting on the
primordial spectrum $P_k$
\begin{equation}\label{eq:rld-start}
C_\ell = \sum\limits_k F_{\ell k} P_k,
\end{equation}
with operator
\begin{equation}\label{eq:operator}
 F^{X Y}_{\ell k} = \Delta \ln k\,  \Delta_{\ell k}^{X} \Delta_{\ell k}^{Y},
\end{equation}
where $\Delta \ln k$ are the logarithmic $k$ intervals for the discrete
sampling chosen in the integration of the system of equations.

A solution for $P_k$ cannot be obtained from a direct inversion of the
$F^{X Y}_{\ell k}$ as it is numerically singular. This is due to the
high level of degeneracy in the transfer functions relating the power
at any wavenumber $k$ to angular multipoles $\ell$. We instead
approximate the inversion by using the SVD method, first reducing the
degeneracy of the system and then inverting using the remaining
orthogonal modes.

The transfer functions can be factorised as
\begin{equation}\label{eq:SVDdecom}
F_{\ell k}= \sum_{\ell'k'}U^{\,}_{\ell \ell'} \Lambda^{\,}_{\ell' k'} V^\dagger_{k' k}\,,
\end{equation}
where the matrices $\bf U$ and $\bf V$ are unitary and of dimensions
$n_\ell$ and $n_k$ respectively, and $\bf \Lambda$ is a non-negative,
diagonal matrix with diagonal elements $\lambda_k$. For this
application, the dimensions of the matrices are that $n_\ell < n_k$
i.e. there are more equations than modes of interest. This results in
some of the diagonal elements of $\bf \Lambda$ being singular
(numerically zero) which prevent the inversion of the transfer matrix.

The SVD method allows one to invert such a system by nulling the
singular modes. This is achieved by creating an inverse $\bf
\Lambda^{-1}$ where the diagonal elements are $1/\lambda_k$ except
where the value of $\lambda_k$ is singular in which case it is
replaced by $0$.

In practice we rank order the factorised modes in descending order of
$\lambda_k$ and all modes with condition number less than a threshold
$\epsilon \mbox{max}(\lambda_k)$ are nulled, $\epsilon \approx 0.038$ for this work. Thus the method is a `$k$-to-$\ell$' compression of
the system where we keep the least degenerate modes connecting the 3d
Fourier space to the 2d angular multipole space. This is not to be
confused with a signal-to-noise compression of the data which aims to
select with respect to orthogonal modes of the covariance of the
observations \cite{Bond:1998qg}.

\begin{figure*}[!htp]
\centering
\includegraphics[width=16.25cm]{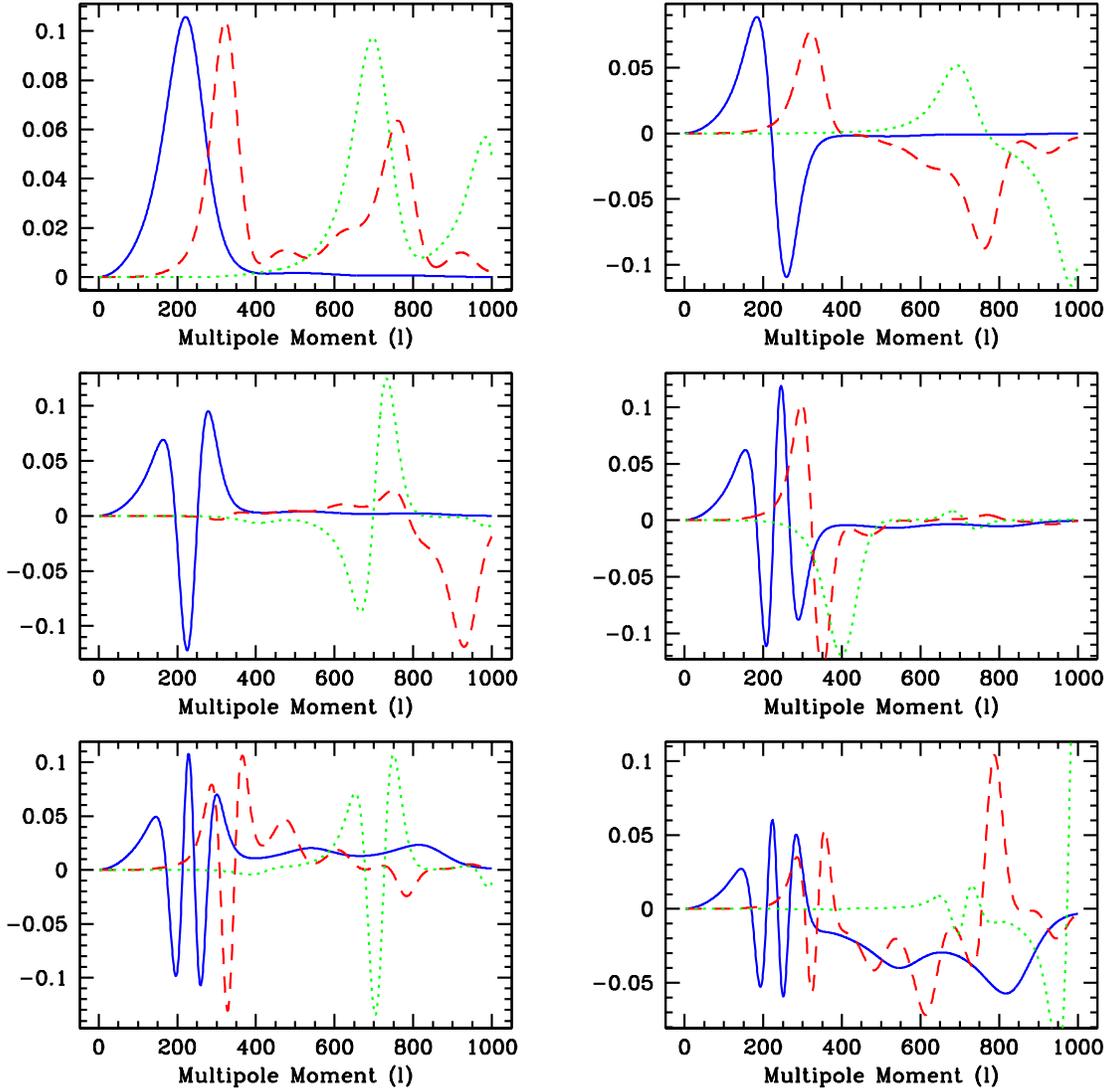}
\caption{We plot the vectors in the matrix $U$ associated with the $6$ highest singular values. The vector corresponding to the highest singular value is shown in the top left panel, the second highest is shown in the top right panel, etc... These vectors are the modes best constrained in $C_\ell$ in the absence of all sources of error. We decomposed $F_{k\ell}$ for a cosmology of $\Omega_bh^2=0.0226$, $\Omega_ch^2=0.108$, $\theta=1.041$ and $\tau=0.076$. The blue/solid line is the $TT$ mode, red/dashed is the $TE$ mode and green/dotted is the $EE$ mode.}
\label{fig:basisvec}
\end{figure*}

It is instructive to look at the first few orthogonal modes given by
the columns of the $\bf U$ matrix. We plot the first six in
Fig.~\ref{fig:basisvec}. These $\ell$-space modes are the least degenerate (or best determined) in the mapping provided by the CMB physics. In other words, in the absence of sample and noise variance, these modes pick out the $\ell$-range where observing the CMB will have the highest impact upon the reconstructed $P(k)$. Not surprisingly, the first few
modes are peaked around the angular scales where the acoustic signal
from each polarisation combination is maximised. As these are the best constrained vectors in
$\ell$ space in the absence of all errors, they are not
necessarily the basis vectors of $P_k$ which are most accurately
constrained, \cite{Hu:2003vp} show what they are for WMAP. We assume
that this ordering of singular values is the optimal method for
sorting the columns in $U$ and $V$.

Once $\bf \Lambda^{-1}$ has been computed the primordial power
spectrum can be reconstructed by inverting a set of observed $C_\ell$ using
\begin{eqnarray}\label{eq:pkrecon}
P_k & = & \sum_{\ell' k' \ell}V^{\,}_{kk'} \Lambda_{k' \ell'}^{-1} U_{\ell' \ell}^\dagger C^{\,}_\ell\,, \nonumber \\
    & \approx & \sum_\ell F_{k\ell}^{-1}C^{\,}_\ell\,.
  \end{eqnarray}
  Our choice of $\epsilon \approx 0.038$ is
  a conservative one with approximately 200 non-singular modes for a
  typical transfer matrix. Reconstructing the primordial power
  spectrum in this way means that we are not using any degenerate
  modes which carry no information in $k$ space but can increase the
  scatter in the reconstructed spectrum. However we are still
  susceptible to the scatter in the observed $C_\ell$ since we have
  not used any noise weighting in this scheme.


  In practice we start our inversion process with a guess input
  spectrum, parameterised by the usual form $A_s k^{n_s-1}$. From this
  and our fiducial cosmological model we obtain a
  $C_\ell^\textrm{model}$ spectrum. We use this to calculate the
  residual spectrum,
\begin{equation}\label{eq:residual}
C_\ell^\textrm{res}=C_\ell^\textrm{obs}-C_\ell^\textrm{model},
\end{equation}
so as to minimise the error induced in $k$-space by the cut-offs in $\ell$, both on large and small scales.

To remove high frequency oscillations in the data we apply a low-pass
filter to the resultant $P_k$. The following algorithm was used,
\begin{equation}\label{eq:lowpass}
P_k^\textrm{low-pass}=\alpha P_k + (1-\alpha)P_{k-1},
\end{equation}
where $\alpha$ was taken to be $0.05$. This method of smoothing leaves
the first few points in the series strongly influenced by
$P_1$. Therefore one should take any effects seen at low $k$ with a
pinch of salt as these points are highly correlated. The covariance
matrix was altered appropriately by a lower-triangular matrix
representing this filter.

We then proceed to bin the reconstructed power spectrum using the
optimal binning method of \cite{Paykari:2009ac}. This binning method
estimates a series of ranges in $k$ over which the signal-to-noise in
the measured primordial power spectrum is constant. Many of the data
points are highly correlated with their nearest neighbours and optimal
binning gives a clear indication of the scales on which we have
independent information. The centre of the $k$ bins chosen by the
cosmological tool {\tt CAMB} \cite{camb} change when the input
cosmological parameters are altered. This is a problem when it is run
over many realisations as in the case of a Markov Chain. We choose the
optimal binning method to find the standardised output of centres and
sizes of $k$ bins for each call of the {\tt CAMB} routine.

To find the optimal binning of the reconstructed power spectrum we
investigate how the uncertainty in the $C_{\ell}$ transfers
into uncertainty in the primordial power spectrum. For this purpose,
we need to define the primordial power spectrum as a series of top-hat
bins:
\begin{equation}
P(k)=\sum_{B}w_{B}(k)Q_{B\:,}
\end{equation}
where $Q_{B}$ is the power in each bin $B$ and $w_{B}=1$ if $k\in B$
and $0$ otherwise. To obtain the errors for these bins we define the Fisher matrix for the $C_{\ell}$ by
\begin{equation}
M_{\ell\ell^{\prime}}=(\delta C_{\ell\ell^{\prime}})^{-1}\:,
\end{equation}
where $\delta C_{\ell\ell^{\prime}}$ is the diagonal matrix of the
squares of the variances in each measurement of $C_{\ell}$. To transfer the given errors from the $C_{\ell}$ to other parameters we
use the Jacobian,
\begin{equation}
M_{\alpha\beta}=\sum_{\ell\ell^{\prime}}M_{\ell\ell^{\prime}}\frac{\partial C_{\ell}}{\partial\theta_{\alpha}}\frac{\partial C_{\ell^{\prime}}}{\partial\theta_{\beta}}\;,
\end{equation}
where, $\theta_{\alpha}$ and $\theta_{\beta}$ represent the bins of the primordial
power spectrum. The derivative of the $C_{\ell}$ with respect
to the primordial power spectrum is the average radiation transfer
function in each bin:
\begin{equation}
\frac{\partial C_{\ell}}{\partial P(k)}=\int_{k_{min}^{B}}^{k_{max}^{B}}\frac{dk}{k}\Delta_{\ell}^{X}(k)\Delta_{\ell}^{Y}(k)\;.\label{eq:Cl_pPS_derivative}
\end{equation}

To calculate the signal-to-noise ratio in each bin we take the inverse square root of the diagonal elements of $M_{\alpha\beta}$
to be the noise and the amplitude of the primordial power
spectrum to be the signal. We then arrange the bins
to have the same signal-to-noise over our $k$ range. Our algorithm will result in more
bins where the signal-to-noise ratio is greater, sampling more finely
where the signal is strongest. 

We construct a signal vector, $S$, which contains the amplitude
of the primordial power spectrum for all the bins and weight our
Fisher matrix by this vector
\begin{equation}
\left(\frac{S}{N}\right)^2_{\alpha\beta}=S_{\alpha}M_{\alpha\beta}S_{\beta}\,,\label{eq:(S/N)^2 FM}
\end{equation}
 where there is no Einstein summation. The square root of the diagonal
elements of this matrix are the $S/N$ of the bins. 

We start our algorithm with the maximum number of bins possible
in our $k$ range. This is set by the usual properties of the Fourier transform.
These imply that the scale of the survey not only determines $k_{min}$,
but also gives a lower bound upon the resolution, $\Delta k_{min}$: narrower bins would become highly
correlated. Therefore, we set up a series of bins with the properties
\begin{equation}
k_{min}=\frac{\ell_{min}}{d_{A}}=\frac{2}{d_{A}}\:\:\:\textnormal{and}\:\:\:
(\Delta k)_{min}=\frac{\Delta\ell}{d_{A}}=\frac{1}{d_{A}}\:,
\end{equation}
where $d_{A}=14.12\textnormal{Gpc}$ (value given by WMAP$5$) is the
angular diameter distance to the surface of last scaterring. We set
$k_{max}=0.08\textnormal{Mpc}{}^{-1}$ as the reconstruction process is
limited past this regime due to the cut-off in $\ell$ at $1000$. A
Fisher matrix is then constructed for this set of bins and weighted by
the signal vector. We choose a signal-to-noise value, that cannot be
less than the maximum value seen in any bin, as our target ratio. The
binwidths are increased in order to reach the target value at each
bin. To obtain this `optimal' binning we iterate until the bin with
the smallest signal-to-noise ratio is within $5\%$ of the target
ratio.

\subsection{Reconstruction with cosmological parameter fitting}

The reconstruction method described above is fast and can be carried
out at each random sample of a MCMC exploration of the cosmological
parameter space. Inserting the reconstruction as part of an MCMC
exploration we can account for the variance induced in the primordial
power spectrum due to the dependence of the radiation transfer
function on the cosmology. 

We have modified the {\tt cosmomc} \cite{cosmomc} package by introducing
the reconstruction at each chain evaluation using the inverse of the
transfer function computed for each combination of parameters acting
on the 'observed' CMB angular power spectrum. The reconstructed
spectrum is then itself used to compute the final $C_\ell$ which are
used to calculate the likelihood at the chain step.

In principle the chains would probe the reduced set of parameters; the
physical densities of baryons $\Omega_bh^2$, and of cold dark matter
$\Omega_ch^2$, the angular diameter distance parameter $\theta$, and
optical depth parameter $\tau$. The primordial power spectrum
parameters $n_s$ and $A_s$ become irrelevant and need not be probed
since the power spectrum is being reconstructed directly. However, in
practice, we {\sl do} include power law spectral parameters which
determined the shape the template model (\ref{eq:residual}) and we
marginalise over the spectral parameters in order to account for any
sensitivity of the reconstruction to the assumed
$C_\ell^{\textrm{model}}$.

The immediate advantage of combining the reconstruction with an MCMC
method is that we can then calculate the variance in the resulting
spectrum due to the random nature of the transfer function. We do this
by including the binned amplitudes for the reconstructed spectrum as
`derived' parameters when analysing the chains. The covariance of the
chains is then mapped into a covariance for the binned power spectrum.

We also need to account for the variance due to the errors in the
observed CMB data. This is not accounted for in the MCMC chains since
we always use the same observed $C_\ell$ data to reconstruct the
spectrum. In principle this contribution to the variance and that
from the transfer function are correlated, however this is difficult
to quantify without including MCMC steps over realisations of the
observations. We therefore make a conservative estimate of the final
error in the reconstructed spectrum by adding the covariance matrix obtained
from the MCMC chain and that obtained by rotating the error matrix of
the observed $C_\ell$ as
\begin{equation}
  \delta P_{BB'} = \sum_{\ell\ell'}F^{-1}_{B\ell}\delta C_{\ell\ell'}F^{-1}_{B'\ell'}\,
\end{equation}
where $F^{-1}_{B\ell}$ is the bin-averaged contribution
from $ F^{-1}_{k\ell}$.

\section{Application of the SVD Inversion}\label{results}

\subsection{Tests on simulated CMB data}

\begin{figure*}[htp]
\centering
\includegraphics[width=14cm]{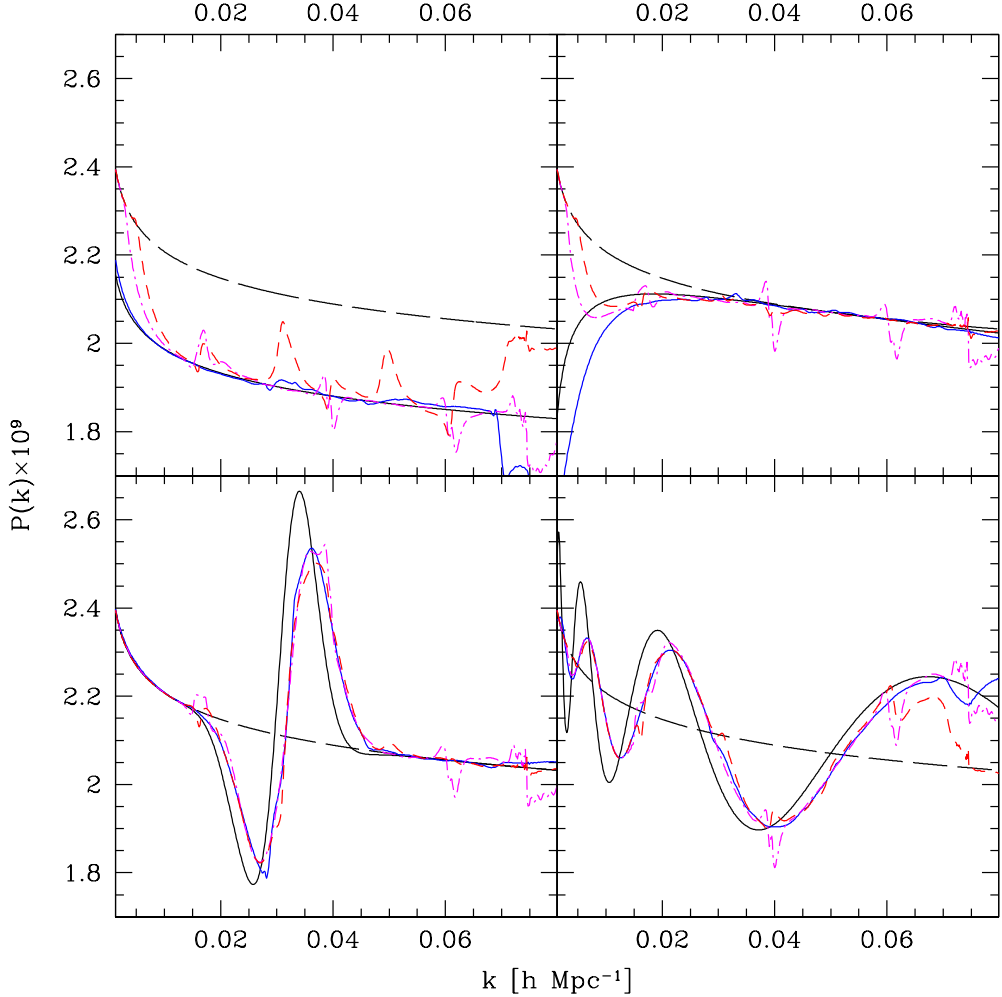}
\caption{The reconstruction of several test spectra. The test models
  used to generate the simulated $C_\ell$, shown in black (thick/solid)
  curves, are (a) A $10$\% decrease in power from the WMAP5
  best fit amplitude, (b) the WMAP5 best fit model including running
  $dn_s/d\ln k = -0.037$, (c) a localised feature at around
  $k=0.02$~Mpc$^{-1}$, and (d) a model with sinusoidal oscillations
  superimposed on the best fit power law spectrum. The black (long-dashed)
  curves show the best fit spectrum used to minimise any cut-off effects at the ends of the $\ell$ regions. The blue (solid) curves are the reconstructions
  using total intensity data whereas the red (dashed) curves and magenta (dot-dashed) curves use
  $TE$ and $EE$ data respectively. The $C_\ell$ forecasts
  assumed an experiment with no noise and an $\ell_\textbf{max} = 1000$.}
\label{fig:testingSVD}
\end{figure*}

\begin{figure*}[htp]
\centering
\includegraphics[width=14cm]{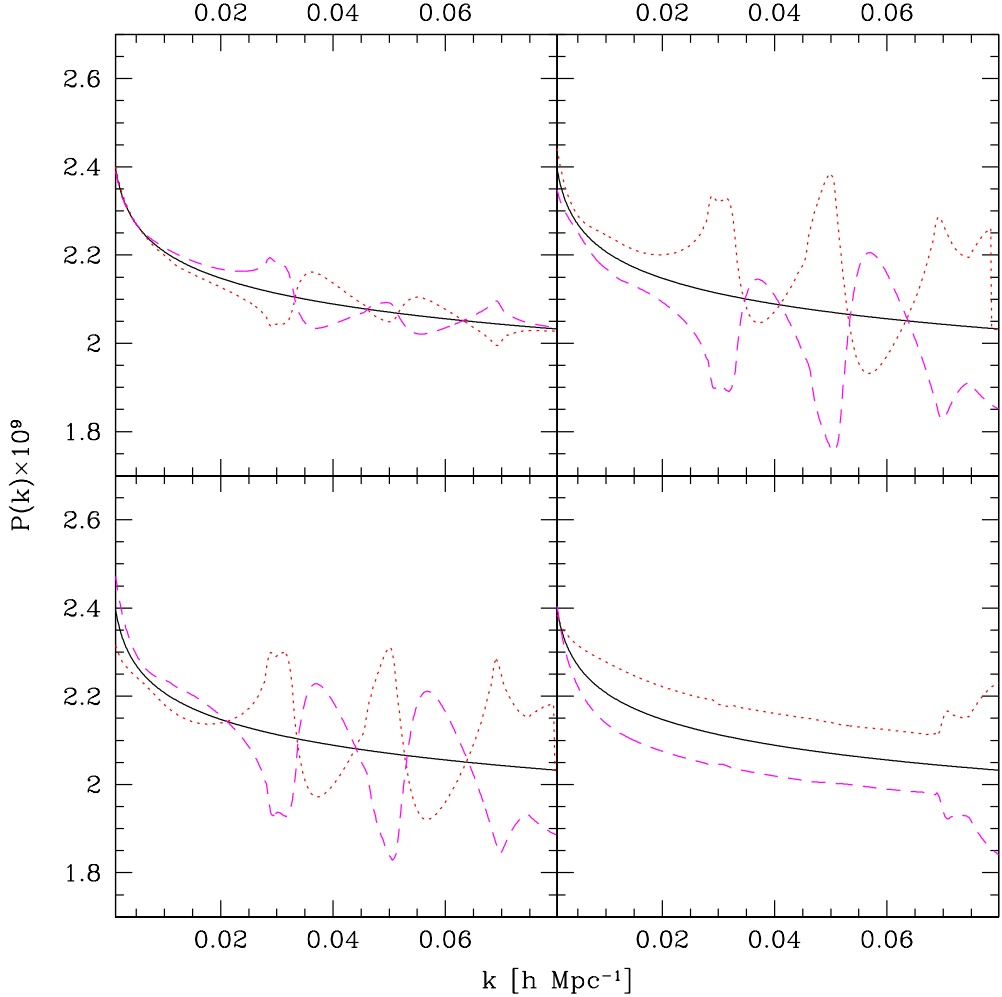}
\caption{The reconstruction of several spectra when the incorrect
  parameters are used for the fiducial cosmological model. The black
  (solid) line is the correct input spectrum. In each panel we change
  a single parameter in the the fiducial model ($\Omega_bh^2=0.0226$,
  $\Omega_ch^2=0.108$, $\theta=1.041$ and $\tau=0.076$). The red
  (dot-dashed) line shows the effect of changing a parameter by
  $1\sigma$ in a positive direction from the WMAP best fit model, the
  magenta (dashed) line shows the effect of a $1\sigma$ shift in a
  negative direction. The top left panel shows the effect of varying
  $\Omega_bh^2$, top right of $\Omega_ch^2$, bottom left of $\theta$
  and the bottom right of $\tau$.}
\label{fig:incorrectparamest}
\end{figure*}

\begin{figure*}[htp]
\centering
\includegraphics[width=14cm]{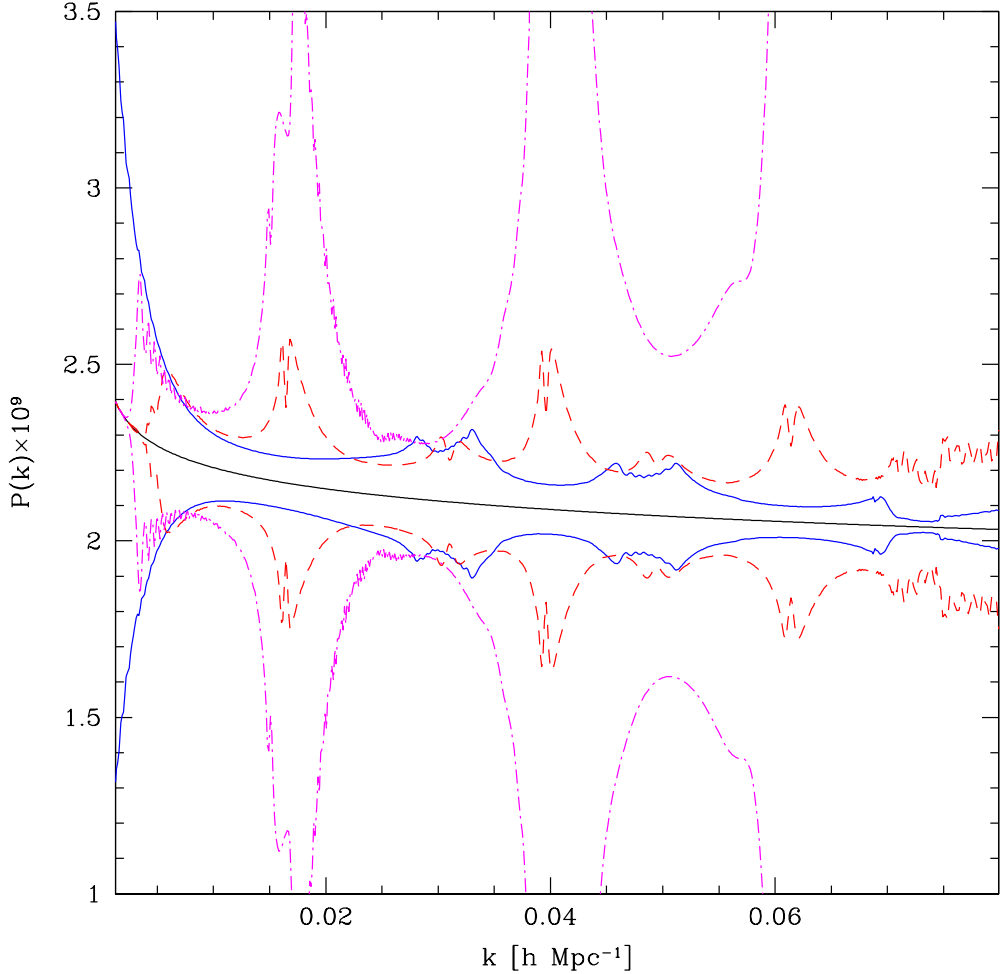}
\caption{We show the unbinned errors on a reconstructed $P(k)$ for a
  forecasted Planck dataset. The cosmological parameters have been
  fixed. The blue (solid) lines show the $1\sigma$ confidence regions
  obtained from $TT$ measurements, with red (dashed) and magenta
  (dot-dashed) showing the same $1\sigma$ bounds for $TE$ and $EE$
  respectively.}
\label{fig:testingSVD-errors}
\end{figure*}

\begin{figure*}[htp]
\centering
\includegraphics[width=14cm]{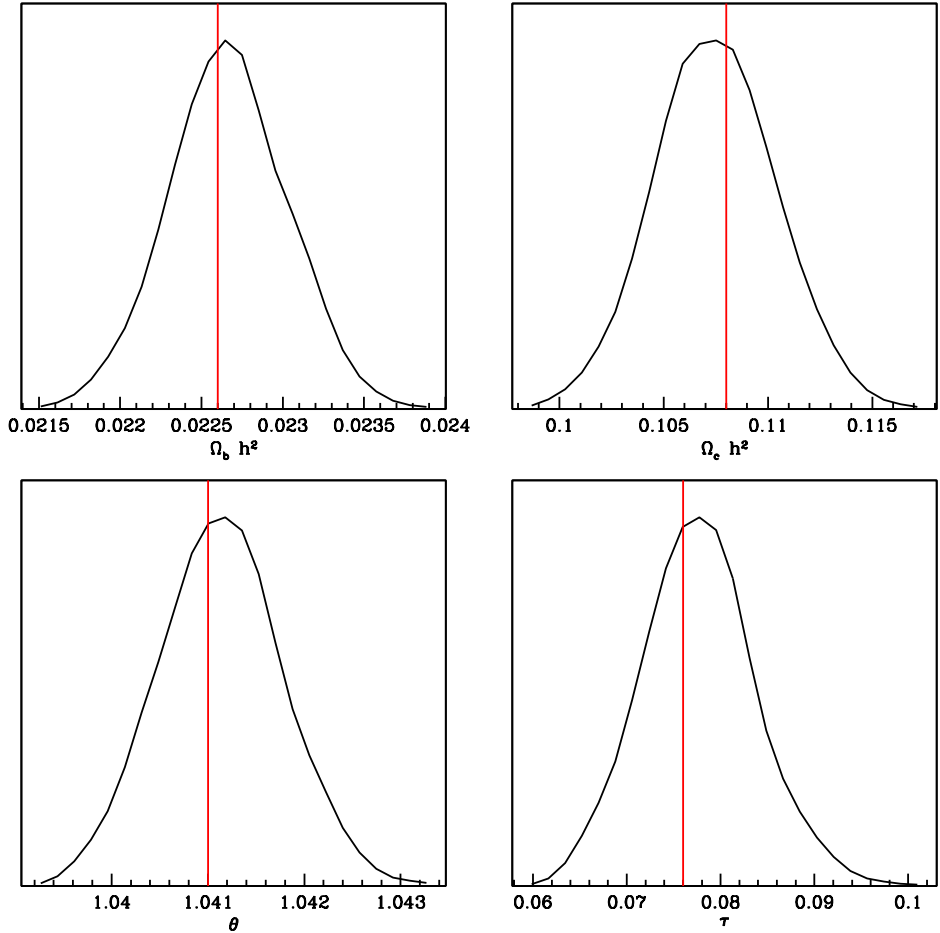}
\caption{Predicted constraints on $\Omega_bh^2$, $\Omega_ch^2$,
  $\theta$ and $\tau$ from Planck, when $P_k$ is given total freedom.
  The solid red vertical lines indicates the input values for each of
  the parameters. The black solid curves show the marginalised
  probability distribution.}
\label{fig:cosmomc-params}
\end{figure*}

\begin{figure*}[htp]
\centering
\includegraphics[width=14cm]{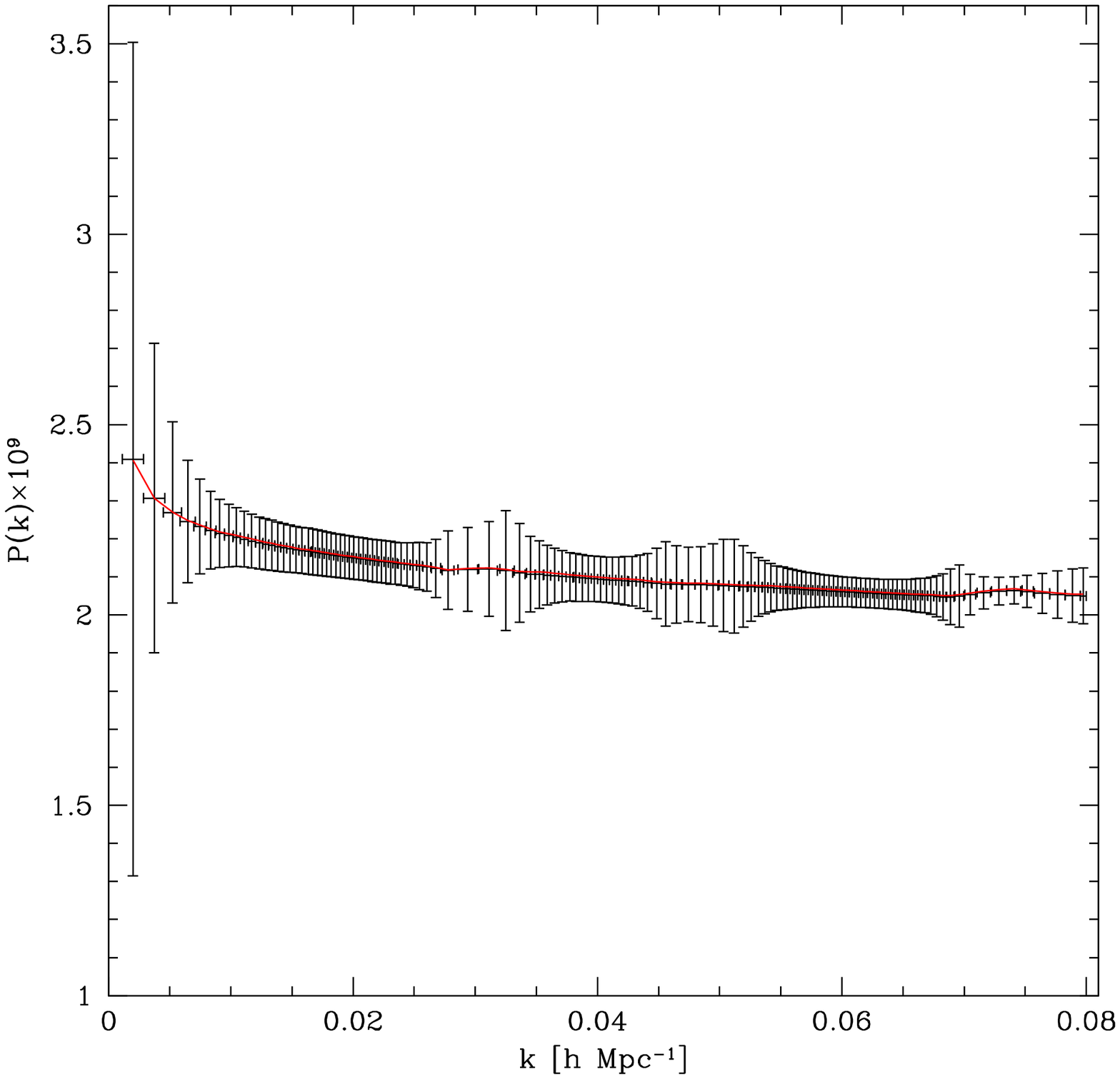}
\caption{Predicted constraints on the form of $P_k$ from Planck. The
  red line is the reconstructed $P_k$ at the best-fit point. We show
  in black the marginalised values of each bin, the error bars
  represent the $1\sigma$ error.}
\label{fig:cosmomc-pk}
\end{figure*}
We start by testing the reconstruction algortihm on a set of input
spectra with known features as in \cite{Nicholson:2009pi}.
Fig.~\ref{fig:testingSVD} shows that the method accurately
reconstructs test features on the input spectra. We used a fiducial
cosmological model of $\Omega_bh^2=0.0226$, $\Omega_ch^2=0.108$,
$\theta=1.041$ and $\tau=0.076$. The starting spectrum used to obtain
$C_\ell$ model is shown by the black, long-dashed line.  We assume
there are no errors on the input $C_\ell^\textrm{obs}$ (between
$\ell=2$ and $\ell=1000$) and observe that the reconstructed $P_k$
generally picks out the input features. We limit our method to fit
between $k=0.0013$ and $k=0.08$ as this is the range with the highest
signal-to-noise in the WMAP data. The features chosen are the same as
in \cite{Nicholson:2009pi} to allow comparisons with this method. They
are a standard power law with $n_s=0.96$ but with an amplitude $90$\%
of that of the best-fit WMAP model, a running $n_s$ model with
$dn_s/d\ln k = -0.037$, a power law with a sharp, compensated feature
at $k=0.02$~Mpc$^{-1}$
\cite{Starobinsky:1992ts,Lesgourgues:1999uc,Adams:2001vc,Feng:2003zua,Mathews:2004vu,Joy:2007na,Hunt:2007dn,Jain:2008dw,Lerner:2008ad}
and a power law with superimposed sinusoidal oscillations
\cite{Wang:2002hf,Martin:2003sg,Martin:2004yi,Pahud:2008ae,Romano:2008rr}. All
four features are clearly recovered to varying degrees when using
$TT$, $TE$ or $EE$, however we find a phase offset between the
reconstructed ad input spectrum for the case where the input spectrum
includes oscillations as in the lower panels of
Fig.~\ref{fig:testingSVD}.  The offset is stable with respect to the
presence of the smoothing kernel, the number of singular values cut
from the inversion, and with repsect to the number of $k$ bins and
range. The reconstructed $P_k$ given by both $TE$ and $EE$ contains
`glitches' not present in the $TT$ reconstruction. These regions
correspond to regions where there is little information int he $TE$
and $EE$ spectra and the reconstruction is still degenerate.

The cosmological parameters obtained from a traditional Markov Chain
Monte Carlo search are used to give us our fiducial operator, $F_{\ell
  k}$. The parameters however have errors present upon them, which are
not usually incorporated into errors on the final reconstructed
$P_k$. If inaccurate parameters have been used to calculate the
fiducial operator very specific signatures would be expected to show
up in the reconstructed from of $P_k$. We show in
Fig.~\ref{fig:incorrectparamest} how these signatures appear in the reconstructed $P_k$ for $TT$ anisotropies
for four cosmological parameters, $\Omega_bh^2$, $\Omega_ch^2$, $h$
and $\tau$. If any features with this form are observered one should
attribute this to an incorrect estimation of the parameters and not
some fundamental physics. A check for this would be if the
corresponding features also show up in the reconstructed $TE$ and $EE$
spectra.

We have also tested the inversion on simulated CMB data with similar
experimental properties as the recently launched Planck satelite mission
\cite{planck}. We assume a total of 12 detectors with NETs of 64$\mu
K/\sqrt{s}$ observing 80\% of the sky over 12 months with a resolution
of 7 arcminutes FWHM. We calculate errors around our fiducial CMB
best-fit models in both total intensity and polarisation spectra for
this experimental setup and use these together with $C_\ell$ samples
on the fiducial model to test the inversion method's properties. We
consider multipoles of $\ell < 1000$ for both total intensity spectra
and polarisation. We have not taken into account any residual error
from foreground subtraction in our forecasts. Thus our forecast are on
the optimistic side of the accuracy achievable in the case of
polarisation where foreground removal will certainly have a
significant impact on errors at $\ell < 1000$. In the case of total
intensity spectra we are significantly underestimating the accuracy
achievable by Planck as we expect to obtain well measured $C_\ell$'s
well above $\ell$ of $1000$. We do not consider these modes as it
significantly increases the time required to perform the
SVD. Consideration of total-intensity modes past $\ell$ of $1000$ will
increase the accuracy obtainable for Planck on $P_k$, it also allows
one to probe $k$ past $0.08$. It would be desirable to perform this
process with a greater $\ell$ range when the Planck data becomes
available.

To compare the accuracy obtainable with each of the anisotropy types
we show the degree to which they each reconstruct a simple power law
$P_k$ in Fig.~\ref{fig:testingSVD-errors}. We use the same fiducial
cosmological model as in Fig.~\ref{fig:testingSVD} with a standard
power law input $P_k$ equal to the spectra being reconstructed. All
the $C_\ell^\textrm{obs}$ were placed on the fiducial model. Over the
whole range of $k$ total-intensity modes best constrain $P_k$. But it
is also true that there are regions in $k$-space, for example between
$k=0.022$ and $k=0.035$, where considering only the $TE$ measurements
can give us a more accurate estimation of $P_k$. $EE$ measurements
approach the accuracy of $TT$ at a number of points, there are however
regions where the errors become so large that any reconstructed $P_k$
is meaningless (these correspond to the troughs of the $EE$
spectrum). The regions of $\ell$-space corresponding to the most
accurately measured $k$ regions are the peaks of the spectra. Both
$TE$ and $EE$ spectra are not significantly affected by changes to
$P_k$ at very low $k$, so errors in this region are artificially
small.

We also tested the combined $P(k)$ and parameter estimation MCMC
search as described in the previous section. The optimal binning
method found $128$ bins for the Planck experiment, where our target
signal-to-noise value in each bin is $10$. We choose this value so as
to have approximately the same number of bins across the range
$0.01<k<0.03$ where WMAP best probes $P_k$.

In Fig.~\ref{fig:cosmomc-params} we show the results of a cosmological
parameter estimation from this process for a simulated Planck
experiment. We find that the input parameters were accurately
recovered after this process. 
We show
the final reconstructed $P_k$ in Fig.~\ref{fig:cosmomc-pk}. The red
line is the reconstructed $P_k$ at its best-fit point. The error bars
we show are obtained from combining the error from the marginalised
distributions with the reconstruction errors given the observed CMB
data. The errors are centred around the mean of the marginalised
distribution for each bin. It is important to note that the errors are
highly correlated. This explains the reduced scatter in the mean
values compared to the size of the plotted errors. As was seen in
Fig.~\ref{fig:incorrectparamest} changing a single parameter by a small amount (in the manner a MCMC search does) creates a very
specific signature on $P_k$ for each parameter, where changing it
slightly has a correlated effect upon the whole range of $k$. This explains the very high correlations observed accross the whole $k$ range. It is
at odds with the errors on $P(k)$ associated with those of $C_\ell$
at any best-fit point which are not correlated across large ranges of
$k$.


\section{Constraints from current CMB observations}\label{current}

\begin{figure*}[t]
\centering
\includegraphics[width=15cm]{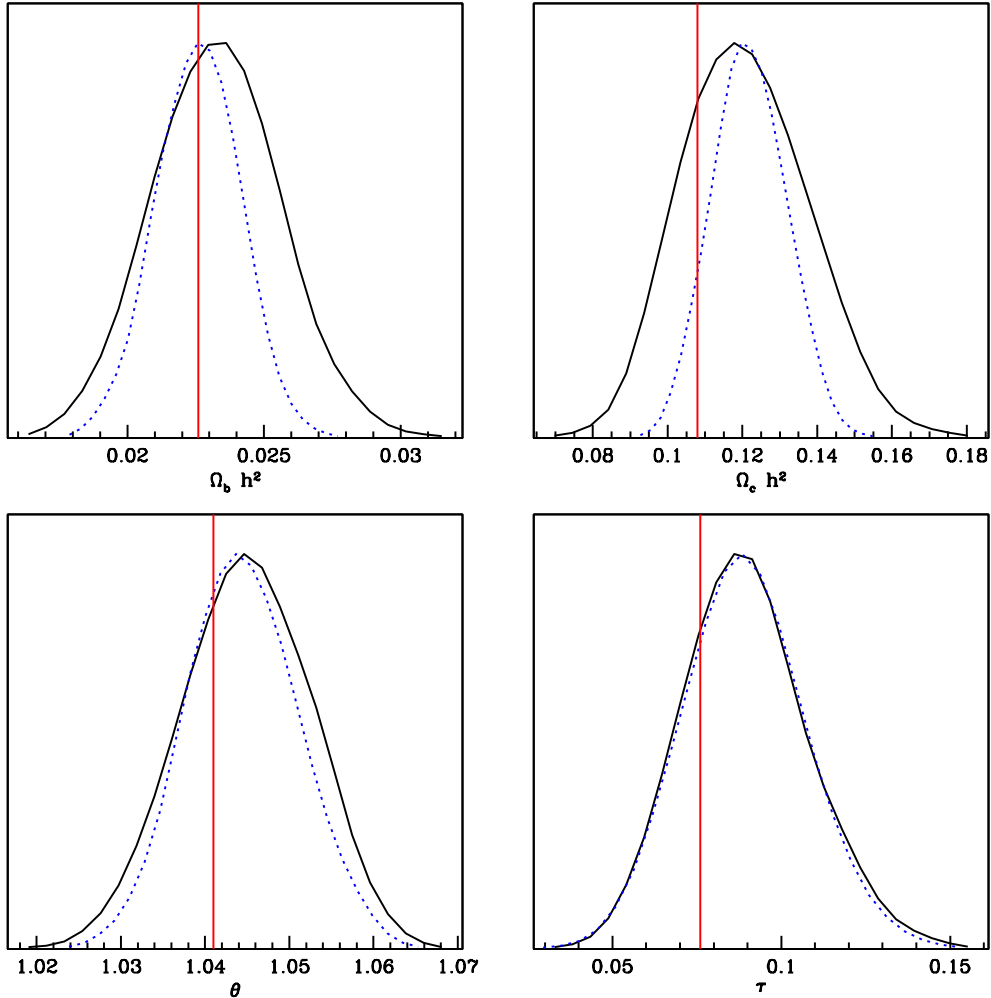}
\caption{The current constraints on $\Omega_bh^2$, $\Omega_ch^2$,
  $\theta$ and $\tau$, when $P_k$ is given total freedom.  The solid
  red vertical lines indicates the WMAP only best fit values when
  $P(k)$ is parameterised in the usual fashion, the black solid curves
  represent the marginalised probability distribution of the WMAP only
  data. The blue dotted line shows the marginalised probability
  distribution of WMAP including the other datasets.}
\label{fig:wmap-pk}
\end{figure*}

\begin{figure*}[t]
\centering
\includegraphics[width=15cm]{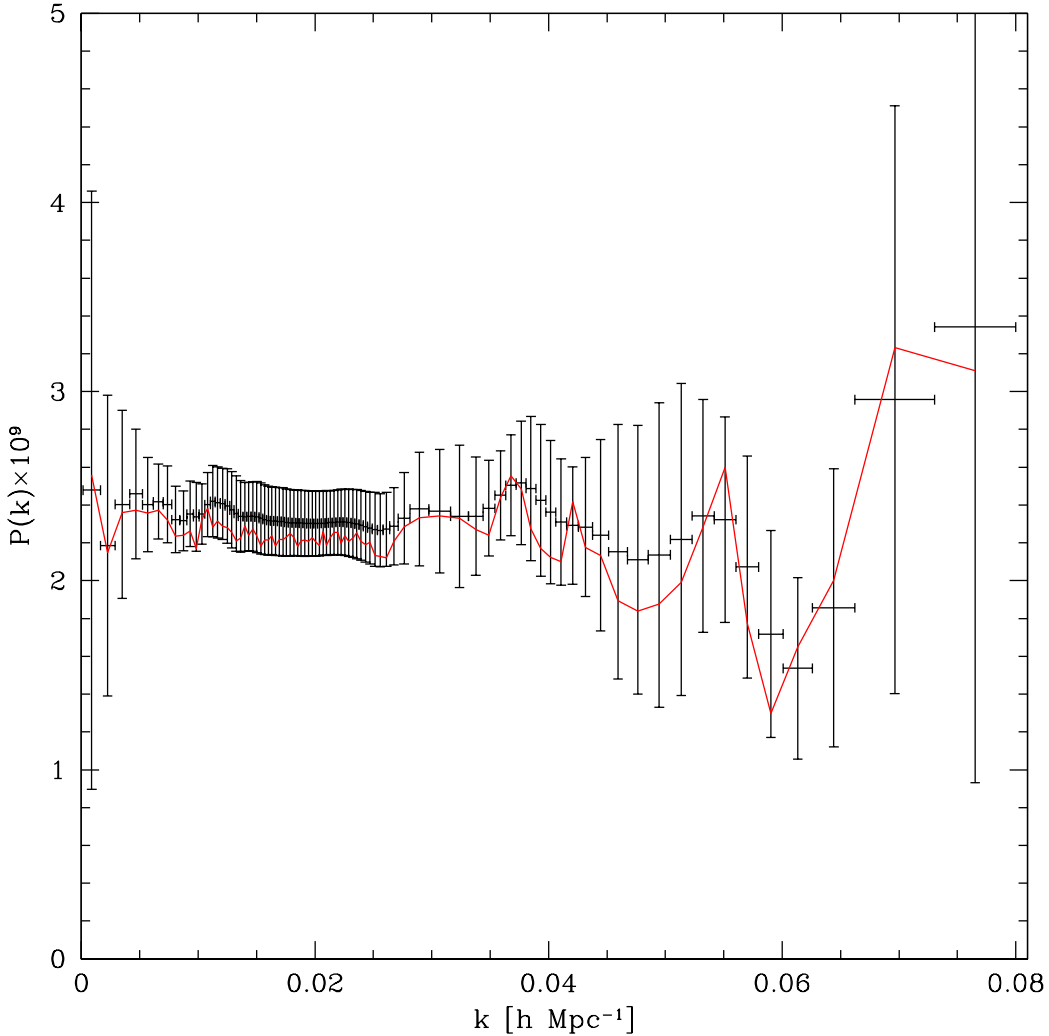}
\caption{The current constraints on the form of $P_k$ from WMAP data
  only. The red lines is the reconstructed $P_k$ at the best-fit
  point. We show in black the marginalised values of each bin, the
  error bars represent the $1\sigma$ error.}
\label{fig:wmap-cosmomc}
\end{figure*}

\begin{figure*}[t]
\centering
\includegraphics[width=15cm]{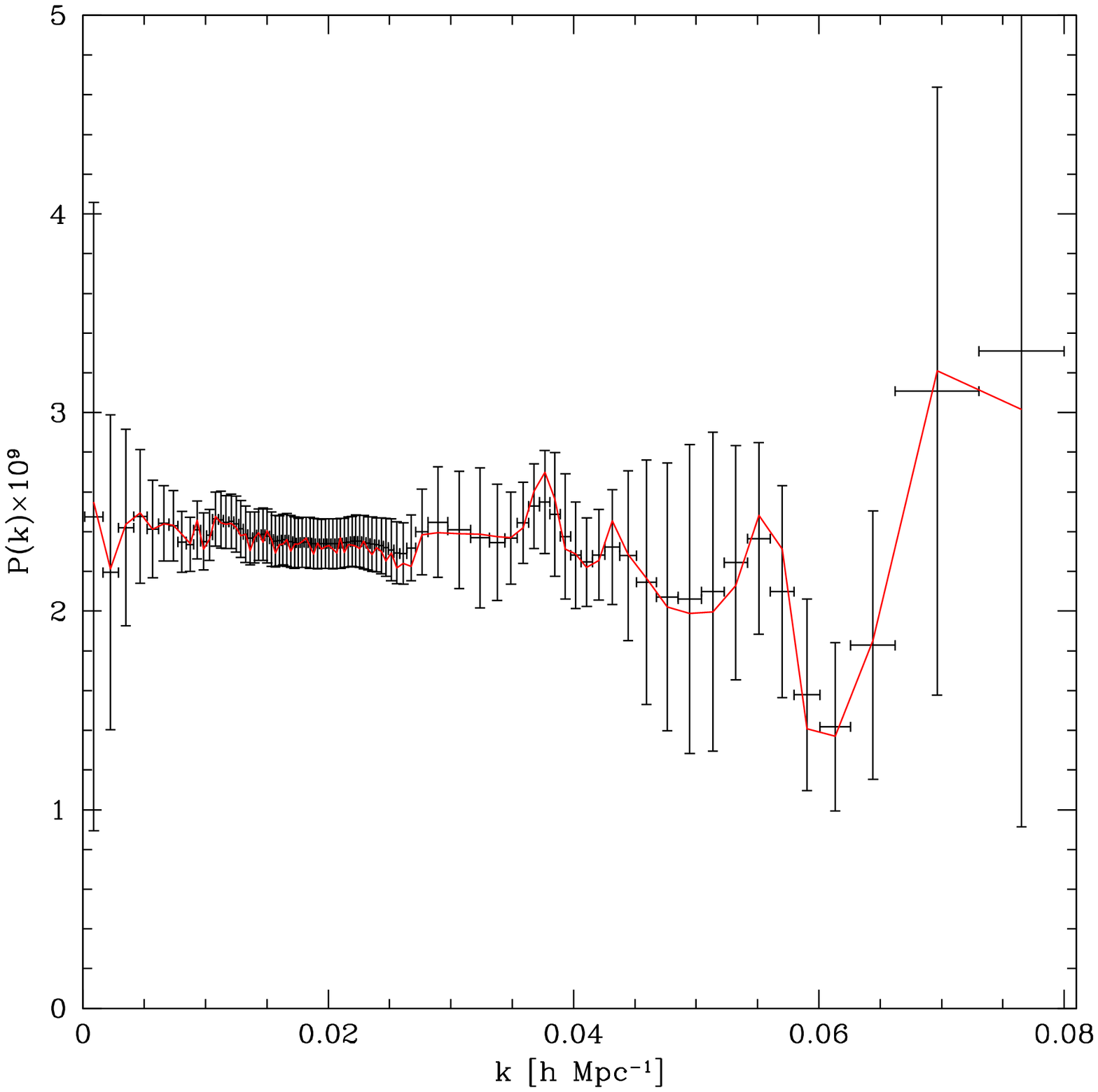}
\caption{The current constraints on the form of $P_k$ from WMAP and
  the other datasets. The red lines is the reconstructed $P_k$ at the
  best-fit point. We show in black the marginalised values of each
  bin, the error bars represent the $1\sigma$ error.}
\label{fig:wmap-cosmomc-sn}
\end{figure*}

We used two sets of currently available data to estimate the
cosmological parameters in conjunction with a free unparameterised
$P_k$. The first is the WMAP 5-year data alone \cite{wmap5}, in the
second we combine this with that of SNIa, HST and BBN
\cite{Perlmutter:1998np,Riess:2004nr,Astier:2005qq,Kowalski:2008ez,Freedman:2000cf}. The
second set combination chosen because the non-CMB sets do not depend
upon the form of $P_k$ and can therefore give us independent and
tighter constraints on most of the cosmological parameters. We run
this data through the MCMC tool {\tt COSMOMC} in the same manner as we
did for the test Planck data.

In Fig.~\ref{fig:wmap-pk} we show the current constraints on
$\Omega_bh^2$, $\Omega_ch^2$, $\theta$ and $\tau$.  The solid red
vertical lines indicates the WMAP only best fit values when $P(k)$ is
parameterised in the usual fashion, the black solid curves represent
the marginalised probability distribution of the WMAP only data. There
is no disagreement between this and the WMAP bestfit model. The blue
dotted line shows the marginalised probability distribution of WMAP
including the other datasets. Here we observe some tension at around
the $1\sigma$ level in $\Omega_ch^2$. The inclusion of this data does
not move the position of the likelihood peak significantly for each
parameter.

The optimal binning method gave us 86 bins in our k range for WMAP. We
used the minimum $\Delta k$ as described previously, this gave a
target signal-to-noise ratio of $6$. The binned reconstructed $P_k$
are shown in Figs.~\ref{fig:wmap-cosmomc} and
\ref{fig:wmap-cosmomc-sn} for each of the two data sets. As in the
Planck case, the results of the MCMC are highly correlated across the
whole $k$ range, in contrast to a single reconstruction which is
not. We have similarly added the errors from the MCMC to a single
reconstruction in each bin in quadrature. The red lines are the
reconstructed $P_k$ at their best-fit points. They error bars are
centred around the best-fit marginalised value of each bin. For the
WMAP only run we find that the limiting errors on $P_k$ in the range
$0.0075 < k < 0.05$ come from uncertainty in the cosmological
parameters, whereas this limiting range is around $0.0075 < k < 0.04$
when we include the other datasets. No significant deviation from the
standard power-law case is observed in either run.

\section{Discussion}\label{discussion}

The SVD based reconstruction method we have outlined provides is fast
enough to be incorporated into full MCMC parameter fitting runs. We
have tested the method and shown that it recovers the overall features
of input spectra. We have applied the method to forecasted Planck data
and current WMAP 5-year results. These results have allowed us to
consistently combine the reconstruction with a full exploration of the
parameter likelihoods for the first time. We have seen that the
limiting factor in constraining the primordial spectrum over a large
range of wavenumbers $k$ comes from the uncertainty in cosmological
parameters. Any claims of a detection of a feature in $P_k$ must
necessarily confront the degeneracy with the cosmological parameter
space . This effect will be less important when the Planck data is
released, however it must still be considered as the unprecedented
accuracy offered by future data may lead to premature claims of a
detection of an interesting feature.

We observed some tension between the WMAP only best-fit model, when
$P(k)$ is parameterised with the normal amplitude and tilt, and our
method in the marginalised probability distribution of
$\Omega_ch^2$. They disagree at around the $1\sigma$ level which is
not overly significant, however it is an indication that there may be
some departure from a standard power law in $P(k)$. Planck
will certainly determine if this is the case.

There are other currently available CMB datasets, which could expand
the range of $k$ probed. However expanding the method to include
multiple data sets is a non-trivial task due to the binning of the
$C_\ell$ data for sub-orbital experiments and the increased running
time required for the expanded $k$-range. We leave this analysis for
future work. 

In the future, as CMB polarisation data becomes increasingly accurate,
it will be desirable to perform a joint inversion of total intensity
data along with polarisation data. It is not clear how to extend the
SVD based method to include all polarisation modes simultaneously
since a HOSVD (Higher-Order SVD) step would probably be required. On
the other hand this would give the best estimate of $P_k$ given any
dataset and would help to reduce the correlations found in the
reconstructed $P_k$ by increasing the degrees of freedom that can be
effectively constrained. 

\acknowledgements

This work was supported by a STFC studentship. We acknowledge the use
the Imperial College high performance computing
service\footnote{http://www.imperial.ac.uk/ict}.

\bibliography{paper}

\end{document}